\documentclass[fleqn,usenatbib]{mnras}

\usepackage[T1]{fontenc}
\usepackage{ae,aecompl}

\usepackage{graphicx}	
\usepackage{amsmath}	
\usepackage{amssymb}	
\usepackage{nicefrac}
\usepackage{multirow}
\usepackage{hyperref}
\usepackage{dcolumn}
\usepackage{siunitx}

\newcommand{\mdot}{\ensuremath{\dot{M}}}
\newcommand\T{\rule{0pt}{2.6ex}}       
\newcommand\B{\rule[-1.2ex]{0pt}{0pt}} 

\newcommand{\Reff}{\ensuremath{R_{\rm eff}}}

\newcommand{\rsun}{\ensuremath{\text{R}_\odot}}
\newcommand{\msun}{\ensuremath{\text{M}_\odot}}
\newcommand{\lsun}{\ensuremath{\text{L}_\odot}}
\newcommand{\logg}{\ensuremath{\log_{10}(g)}}
\newcommand{\logge}{\ensuremath{\log_{10}(g_{\rm e})}}
\newcommand{\loggp}{\ensuremath{\log_{10}(g_{\rm p})}}
\newcommand{\loggl}{\ensuremath{\log_{10}(g^\ell)}}
\newcommand{\gloc}{\ensuremath{g^\ell}}
\newcommand{\loggN}{\ensuremath{\log_{10}(g_{\rm N})}}
\newcommand{\logL}{\ensuremath{\log_{10}\left({L/\lsun}\right)}}

\newcommand{\sini}{\ensuremath{\sin{i}}}
\newcommand{\vesini}{\ensuremath{v_{\rm e}\sin{i}}}
\newcommand{\veq}{\ensuremath{v_{\rm e}}}

\newcommand{\kms} {\mbox{\rm km$\;$s$^{-1}$}}

\newcommand{\teff}{\ensuremath{T_{\rm eff}}}
\newcommand{\teffl}{\ensuremath{T^{\ell}_{\rm eff}}}
\newcommand{\rpole}{\ensuremath{R_{\rm p}}}
\newcommand{\reqtr}{\ensuremath{R_{\rm e}}}
\newcommand{\prot}{\ensuremath{P_{\rm rot}}}
\newcommand{\pphot}{\ensuremath{P_{\rm phot}}}

\newcommand{\omomc}{\ensuremath{\omega_{\rm e}/\omega_{\rm crit}}}
\newcommand{\omcrit}{\ensuremath{\omega_{\rm crit}}}

\newcommand{\gaia}{\textit{Gaia}}
\newcommand{\hipp}{\textit{Hipparcos}}
\newcommand{\sqigl}{\raisebox{1pt}{\rotatebox[origin=c]{75}{$\sim$}}}

\newcommand{\HeII}{\mbox{He\,\textsc{ii}}}
\title[Distance, rotation, and parameters of $\zeta$~Pup]{The distance, rotation,
and physical parameters
of $\zeta$~Pup}

\author[Ian D. Howarth \& Floor van Leeuwen]{
Ian D. Howarth$^1${\thanks{e-mail: i.howarth@ucl.ac.uk}}
and Floor van Leeuwen$^2${\thanks{fvl@ast.cam.ac.uk}}
\\
$^1$Department of Physics and Astronomy, University College London,
Gower Street, London WC1E 6BT, UK\\
$^2$Institute of Astronomy,
University of Cambridge,
Madingley Road,
Cambridge
CB3 0HA,
UK
}

\date{Accepted VVV. Received YYY; in original form ZZZ}

\pubyear{2018}

\begin{document}
\label{firstpage}
\pagerange{\pageref{firstpage}--\pageref{lastpage}}
\maketitle

\begin{abstract}
  We scrutinize the \hipp\ parallax for the bright O super\-giant
  $\zeta$~Pup, and confirm that the implied distance of $332 \pm
  11$~pc appears to be reliable.  We then review the implications for
  the star's physical parameters, and the consequences for the
  interpretation of \pphot, the \mbox{1.78-d} photo\-metric period.
  The equatorial rotation period is $<$3.7~d (with 95\%\ confidence),
  ruling out a proposed $\sim$5.1-d value.  If the photometric period
  is the rotation period then $i$, the inclination of the rotation
  axis to the line of sight, is $33{\fdg}2 \pm 1{\fdg}8$.  The
  inferred mass, radius, and luminosity are securely established to be
  less than canonical values for the spectral type, and are not in
  agreement with single-star evolution models.  The runaway status,
  rapid rotation, and anomalous physical properties are all indicative
  of an evolutionary history involving binary (or multiple-star)
  interaction.  We perform simple starspot modelling to show that the
  low axial inclination required if $\prot = 1.78$~d has testable
  spectroscopic consequences, which have not been identified in
  existing time series.  If \pphot\ is directly related to
  drivers of systematic, high-velocity stellar-wind variability 
  (`discrete absorption components') in $\zeta$~Pup, antisolar differential rotation
  is required.  Model line profiles calculated on that basis are at variance
  with observations.

\end{abstract}

\section{Introduction}

As the brightest early-O super\-giant by a comfortable margin,
$\zeta$~Pup (HD~66811, O4~I\,(n)fp; \citealt{Sota14})
has long been a touchstone in the development of models of
radiatively-driven winds, and their role in massive-star evolution.
In parallel, it has been subject to observational scrutiny at all
accessible wavelengths, from X-ray to radio (e.g., \citealt{Blomme03,
  Eversberg98, Hanson05, Harries96, Marcolino17, Naze18, Reid96}; and
many others).  However, it was only with the relatively recent
availability of satellite photo\-metry that an apparently periodic signal with
$\pphot = 1.78$~d was discovered in its optical brightness
\citep{Howarth14, TahinaR18}.

The origin of this signal remains moot. In a major study
built on photometry obtained with 
\textit{BRITE-Constellation} nanosatellites \citep{Weiss14}, \citet{TahinaR18} argued for it
to be the rotation period, an interpretation that \citet{Howarth14}
had considered less likely than a pulsational origin,
on the grounds that exceptional, near-critical
rotation would be implied -- the projected equatorial rotation
velocity of $\zeta$~Pup, $\vesini \simeq 213$~\kms\
($\S$\ref{sec:naive}), is already the most rapid known for any Galactic
O~super\-giant \citep{Howarth97}, and a \mbox{1.78-d} rotation period would
require that \veq\ be $\sim{2}\times$ greater.  Nevertheless, this
is merely a plausibility argument, and exceptionally rapid rotation
could simply be a signature of an exceptional evolutionary history
(cf., e.g., \citealt{Vanbeveren12, deMink13}).

The case of $\zeta$~Pup therefore represents a modern manifestation of
the long-standing difficulty in making a compelling distinction,
observationally, between pulsational and rotation modulation as the
mechanism responsible for low-amplitude spectroscopic and
photo\-metric variability in early-type stars (cp., e.g.,
\citealt{Gies91, Harmanec99}).  This problem has been compounded in
the case of $\zeta$~Pup by some contention in respect of its distance,
and consequently in basic physical parameters such as radius and mass.
Our purpose in this paper is to examine these issues.

To do so, we first review the \hipp\ parallax data in
Section~\ref{sec:dist} (and examine the \hipp\ photometry in
Section~\ref{hipp:phot}).  We evaluate basic stellar parameters in
Section~\ref{sec:params}, under the limiting assumptions of
(\textit{i}) axial inclination $i = 90^\circ$ and (\textit{ii})
rotation period $\prot = \pphot (= 1.78$~d).  Evolutionary implications are
discussed in Section~\ref{sec:disco}.  Physical modelling of
photometric and spectroscopic variability, intended to test the
rotational-modulation/hotspots hypothesis, is presented in
Section~\ref{sec:Spots}.  The case for an association between \pphot\
and discrete absorption components in the stellar wind is scrutinized
in Section~\ref{sec:cirdac}.  Finally, the summary and conclusions are
given in Section~\ref{sec:summ}.

\section{Distance}
\label{sec:dist}

%
%

\subsection{\hipp}
\label{sec:hpllx}

Although the revised reduction of the \hipp\
data yields a reasonably precise parallax of $(3.01 \pm 0.10)$~mas
for $\zeta$~Pup (distance $d = 332 \pm 11$~pc; 
%
\citealt{vanLeeuwen07a, vanLeeuwen07b}), subsequent state-of-the-art
model-atmosphere analyses have disregarded or challenged this result
(which implies unexpectedly low values for the stellar mass and
radius,\footnote{Equivalently, the implied absolute magnitude, $M(V) =
-5.5$ ($\S$\ref{sec:prelim}),
is
$\sim$1$^{\rm m}$ fainter than canonical values for the spectral type (e.g.,
\citealt{Walborn73, Martins06}), and is close to expectations for a
main-sequence star.}
for standard evolutionary scenarios), preferring larger distances (up
to $\sim$700~pc; e.g., \citealt{Najarro11, Bouret12, Pauldrach12}), with
concomitant implications for the luminosity, etc.

Unfortunately, at $V \simeq 2.1$ $\zeta$~Pup is too bright to have
been included in currently available \gaia\ data releases.  However,
its brightness is an asset where \hipp\ is concerned
(parallax errors are limited by photon noise at $V \sim 3$ and fainter,
although by calibration uncertainties otherwise\footnote{Compared to the
  original analysis, calibration
  uncertainties were reduced by a factor $\sim$5 in the 2007 re-reduction.}),
and we have reviewed the
results to check if there are any reasons to
suspect the published parallax.

We find no suggestion of any problems in the astro\-metric data; the
error correlations for the astro\-metric parameters are very low, and
the distribution over scan directions very good, as is the
distribution over parallax factor.  The underlying data for the star
are consistent and numerous (138 observations, with only 4 rejections
in the iterative solution); and $\zeta$~Pup is in a part of the sky where
the scan coverage is almost maximally good.

[The parallax factor multiplied by the actual parallax of the star
gives the along-scan displacement of the position due to the parallax
at the time of observation, for the given scan direction; it always
lies in the range $-0.7$:+0.7. For $\zeta$~Pup the parallax factors
fall entirely in the ranges $-0.7$:$-0.4$ and +0.4:+0.7, which
is very good distribution for a reliable parallax determination.  The spread over
epochs is also very good, giving a low ($\sim\mathcal{O}$(0.1)) error
correlation between proper-motion and parallax determinations.]
 
\begin{figure}
\centering
\includegraphics[scale=0.5,angle=270]{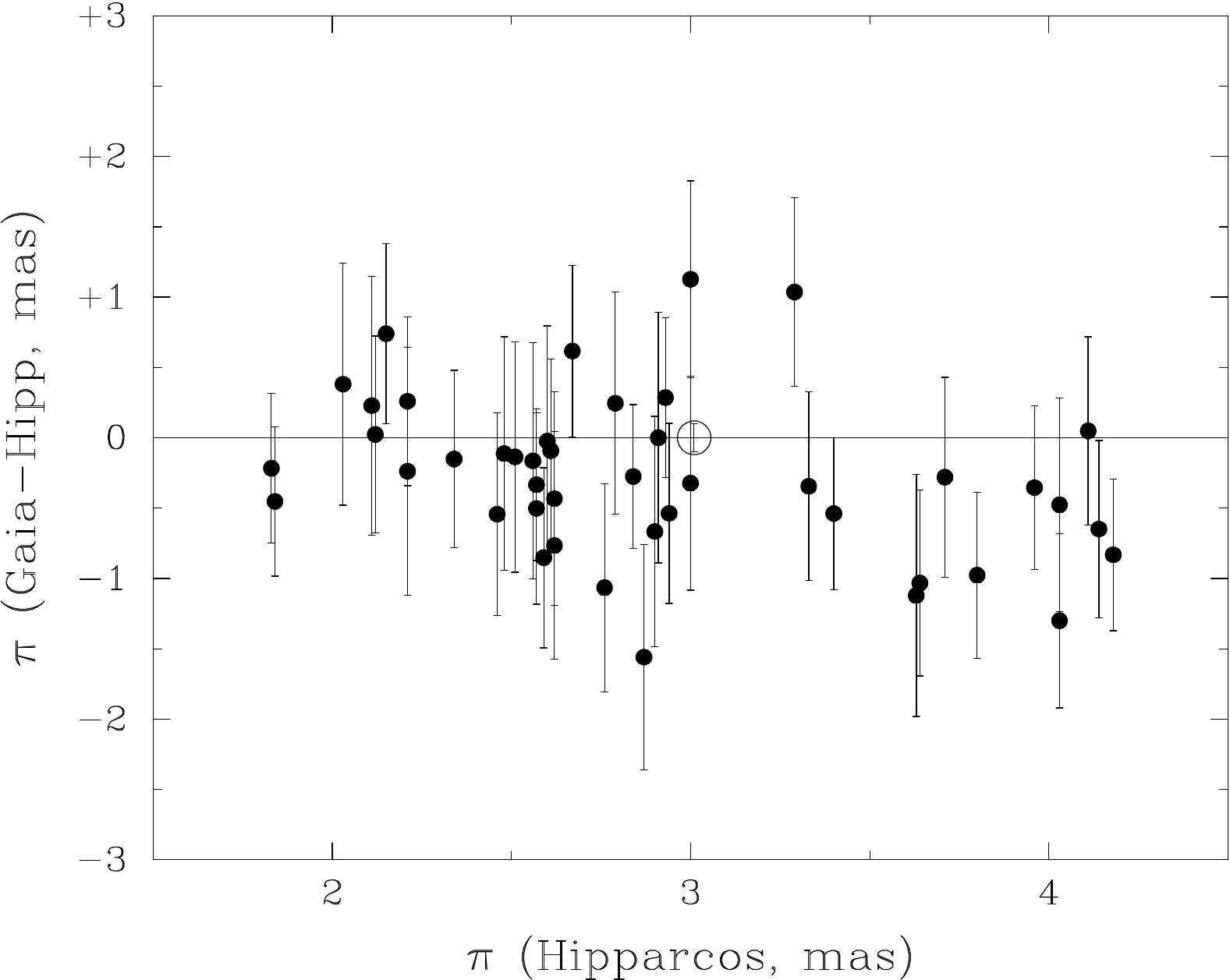}
\caption{Differences between \hipp\ and \gaia\ parallaxes for stars
  within 3$^\circ$ of $\zeta$~Pup, as a function of \hipp\
  parallax.  Measurements with less than 3$\sigma$ significance are
  excluded, as are those with \hipp\ (\gaia) errors greater than~1
  (0.1)~mas.   The \hipp\ $\zeta$~Pup measurement is shown as an
  open circle, with radius equal to the 1-sigma error (which is
  smaller than for other stars shown because of $\zeta$~Pup's brightness).}
\label{fig:HipG}
\end{figure}

As a further check, we compared results for apparently single stars
within
%
%
180\arcmin\ of $\zeta$~Pup that have measurements both from
\hipp\ and in \gaia\ DR2 \citep{Gaia18a}.   Fig.~\ref{fig:HipG} shows the
result.   This confirms that there are no reasons to
doubt the \hipp\ measurement.

Finally, given that 
the parallactic displacement 
is $\sim$10$\times$ greater than the star's
angular diameter
(Section~\ref{sec:diam}), and that the
amplitude of photo\-metric variability is small, any
asymmetry in the surface-brightness distribution is very unlikely
to compromise this conclusion.

\subsection{Corroboration}

Since the discovery of the Gum nebula, its ionization has generally
been attributed to $\gamma^2$~Vel and $\zeta$~Pup \citep{Gum52}.
Prior to the advent of satellite astro\-metry, the distance to
$\zeta$~Pup was therefore estimated on that basis (e.g., ``both
$\gamma^2$~Vel and $\zeta$~Pup appear to be embedded in the giant Gum
{H\,\textsc{ii}} region and are the sources of its ionization so that
we can assume the two stars are at the same distance'';
\citealt{Morton69}).

In support of this assumption, \citet{Woermann01} argued that the Gum
nebula could be the remnant of a super\-nova from a previous binary
companion to $\zeta$~Pup, noting that the surviving O~star passed within
$\nicefrac{1}{2}^{\circ}$ of the expansion centre of the nebula about
1.5~Myr ago.  They further concluded that the nebula is probably
primarily ionized by $\zeta$~Pup, at a distance in the range $\sim$200--500~pc.

A physical association of the various components of the `Vela
Complex', including the Vela super\-nova remnant, $\gamma^2$~Vel and
$\zeta$~Pup, the Gum nebula, and the Vela~OB2 association, has also
been widely assumed (e.g., \citealt{Sushch11}).  The significance of
this is that apparently reliable distant estimates for other
components of the Vela Complex can be used as a check on the
plausibility of the \hipp\ parallax for $\zeta$~Pup.

The $\gamma^2$~Vel binary system is particularly useful in this
context, as its distance can be independently established by primary
(geometric) means.  \citet{Millour07} obtained a distance $d =
368^{+38}_{-13}$~pc by combining new interferometric separation
measurements with spectroscopic-orbit observations, a result
independently confirmed and refined by \citeauthor{North07}
(\citeyear{North07}; cf.\ also \citealt{Lamberts17}), who, in effect,
solved the orbit in three dimensions to obtain $d = 336^{+8}_{-7}$~pc.
The new reduction of \hipp\ data for $\gamma^2$~Vel gives a distance
$d = 342^{+40}_{-32}$~pc \citep{vanLeeuwen07a, vanLeeuwen07b}, in
excellent agreement.  Further corroboration is provided by a precise
photometric determination of the distance to Vela OB2, yielding $d = 350\pm 13$~pc
\citep{Jeffries09}.

We
conclude that independent determinations of the distances to Vela~OB2
and to $\gamma^2$~Vel are in good mutual agreement, and both are in
good accord with the \hipp\ distance to $\zeta$~Pup.  Once again,
there appear to be no good grounds to doubt the reliability of the
\hipp\ parallax for $\zeta$~Pup.

\subsection{The runaway $\zeta$ Pup}

That $\zeta$ Pup is a runaway star was first proposed by
\citet{Upton71}, and its dynamics and origin have subsequently been
discussed a number of times (e.g., \citealt{Blaauw93, vanRensbergen96,
Moffat98,
Hoogerwerf01, Schilbach08}).

The \hipp\  parallax and proper motion ($34.07 \pm
0.10$~mas~yr$^{-1}$) yield
a transverse velocity of $53.7 \pm 1.8$~\kms.  We found four primary
literature sources that yield 19 separate radial-velocity measurements
(\citealt{Frost26, Wilson63, ContiLL77, Garmany80});
%
%
%
%
those measurements are in satisfactory mutual agreement, with roughly
similar estimated errors (and provide no evidence for
binarity). We adopt their unweighted average, $-25.4
\pm 2.1$~\kms\ (s.e.).

The space velocity with respect to the Sun is thus $59.4\pm 1.9$~kms.
Correcting for the Sun's peculiar motion and for Galactic
rotation\footnote{Using (U,V,W)$_\odot = (11.10, 12.24, 7.25)~\kms$
  \citep{Schonrich10}; $R_0 = 8.5$~kpc, $\theta_0 = 220~\kms$
  \citep{Kerr86}; and the Galactic pole position adopted in
  \citet{ESA97}.  Quoted errors do not include uncertainties on these
  quantities.} we obtain a peculiar velocity of $56.2 \pm 1.9~\kms$
(transverse and radial components 36.6, $-42.7$~\kms).  These figures
confirm that $\zeta$~Pup is a runaway by any generally accepted
definition (e.g., \citealt{Gies87}).

However, the \hipp\ distance rules out the runaway scenarios discussed
by \citeauthor{vanRensbergen96} (as already noted by
\citealt{Schilbach08}), and hence also the \textit{specific}
binary-merger evolutionary scenario proposed by \citet{Vanbeveren12}
and discussed by \citet{Pauldrach12}.  The most likely site of origin
for $\zeta$~Pup appears to be the cluster Trumpler~10
\citep{Hoogerwerf01, Schilbach08} on the basis of its \hipp\ distance
($d = 386 \pm 5$~pc; \citealt{vanLeeuwen09}), although the
\textit{Gaia} value ($441\pm 4$~pc; \citealt{Gaia18}), based on a much
larger sample of stars, may require a review of this conclusion.

\begin{figure}
\centering
\includegraphics[scale=0.43,angle=0]{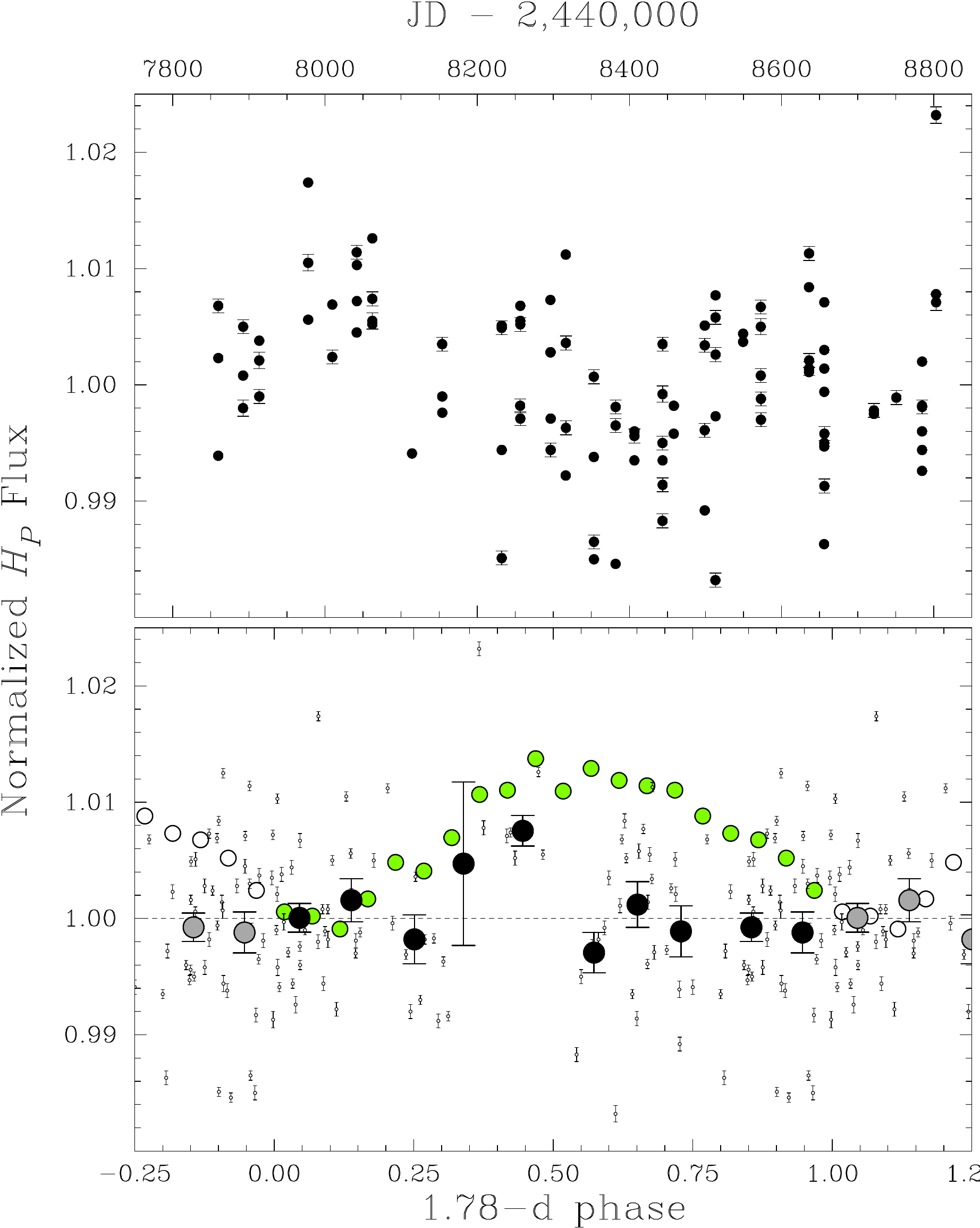}
\caption{Upper panel: \hipp\ epoch photo\-metry of $\zeta$~Pup
  (excluding those with error bars $>0.001$), discussed in
  Section~\ref{hipp:phot}.  \newline Lower panel: phase-folded data.
  Small dots are individual \hipp\ observations; large dots with
  error bars are unweighted means in 0.1-phase bins. 
For reference,
  \textit{SMEI} results from \citet{Howarth14} are shown in green;
  error bars are slightly smaller than symbol size.}
\label{fig:HipLC}
\end{figure}

\section{\textit{Hipparcos} photo\-metry}
\label{hipp:phot}

As well as astro\-metry, \hipp\ provided relatively precise
broad-band photo\-metry;  results for $\zeta$~Pup are shown in
Fig.~\ref{fig:HipLC}.  The formal errors may not be reliable for such
a bright target, due to saturation effects, but nevertheless the
dispersion of the data (s.d. 7.5~mmag) is consistent with the
stochastic microvariability on $\sim$10-hr timescales reported by
\citeauthor{TahinaR18} (\citeyear{TahinaR18}; see also
\citealt{Balona92}), with  possible longer-term changes at the
$\sim$1\%\ level.

These data were previously examined by \citet{Marchenko98}, who found
a 2.563-d periodicity with semiamplitude 6~mmag (again,
tentatively attributed to rotational modulation).  Our time-series
analysis of these data recovers this result, but shows no evidence for
a significant periodic signal close to $\pphot=1.78$~d.
Fig.~\ref{fig:HipLC} includes the photo\-metry phase-folded on an
ephemeris with $T_0 = \text{JD }2\ 448\ 464.0$ (close to the mean date
of observation), $P = 1.780\ 938$~d \citep{Howarth14}, and binned in
phase intervals of 0.1.  Error bars shown on the binned points are
standard errors, using standard deviations in each bin computed from
the dispersions of the data (and not the nominal errors on individual
points).

These binned data are not formally consistent with a phase-constant
flux value, with $\chi^2 = 38.7$, but this rather large value is
dominated by a single outlying bin, at $\phi \simeq 0.45$ in
Fig.~\ref{fig:HipLC}. Of only five observations falling in that bin,
four accordant results come from a single epoch; the 5th observation
is well separated in time, but differs by less than 1~mmag from the
mean brightness, resulting in a very small error bar.  We conclude
that the phased data are consistent with a near-constant time-averaged
flux (semi-amplitude $\lesssim$3~mmag) over at least 80\%\ of the
\mbox{1.78-d} period.

A \mbox{1.78-d} periodic signal in the photo\-metry could be smeared
out by phase drift.  However, examination of subsets of the epoch
photometry shows no evidence for short-lived, large-amplitude periodic
signals.  Moreover, the \textit{SMEI} discovery data have a very
similar time-span to the
\hipp\ mission ($\sim$1000~d), yet show obvious, near-coherent,
$\sim$sinusoidal changes (Fig.~\ref{fig:HipLC}).  Additionally,
\citet{Balona92} failed to detect a signal in precise ground-based
photo\-metry obtained in 1989 April, $\sim$four months before the
first \hipp\ observations.  A plausible interpretation is that, around
$\sim$1990, the \mbox{1.78-d} signal was of significantly lower amplitude than
in $\sim$2005 and 2015, when a $\gtrsim$5~mmag semi-amplitude was
recorded \citep{Howarth14, TahinaR18}.

\section{Rotation and physical parameters}
\label{sec:params}

Given the apparently well-determined distance, we can 
review the implied physical parameters for $\zeta$~Pup, 
including its rotation (with the ultimate aim of testing the
hypothesis that the \mbox{1.78-d} photo\-metric signal is a direct tracer of
the rotation period).

\subsection{Principles \& preliminaries: slow-rotation limit}
\label{sec:prelim}

The basic principles are simple: the observed and surface fluxes lead
to an estimate of the angular diameter; the angular diameter and
distance give the radius; the radius and an assumed rotation period
lead to the equatorial rotation speed, {\veq}; which, with the
observed projected rotation speed, gives the inclination of the
rotation axis to the line of sight,~$i$.  The radius, effective
temperature, and surface gravity also yield the luminosity and
(`spectroscopic') mass.  

These are straightforward sums as long as the
star is satisfactorily approximated as a sphere of uniform surface
flux.  This is marginal for $\zeta$~Pup; its rapid
rotation introduces complications that are considered further in
Section~\ref{sec:prac}.  Nevertheless, an examination of parameters in the
spherical-star, or slow-rotation, limit is of use to provide a context
and a point of comparison with previous analyses, and allows us to
assemble some necessary numerical data.

\subsubsection{Observed flux} The observed $V$ magnitude and colours
yield the extinction.  We took $V = 2.25$, $(B-J) = -0.82$
\citep{Johnson66} to estimate $E(B-V)=0.04$, and thence a
reddening-corrected visual magnitude $V_0 = 2.13$.  Using the \hipp\
parallax, the implied absolute magnitude is $M_V = -5.48 \pm
0.09$.

The corresponding flux is
\begin{align*}
f_V = (5.14 \pm 0.17) &\times 10^{-10}\text{~erg cm$^{-2}$ s$^{-1}$  \AA$^{-1}$}\\
[\equiv (5.14 \pm 0.17) &\times 10^{-12}\text{~J m$^{-2}$ s$^{-1}$  nm$^{-1}$}].
\end{align*} Here we have used an extinction law, intrinsic colours,
and flux calibration from \citet{Howarth83, Howarth11a}.  The error
quoted on the observed flux is based on 1\%\ uncertainties
in each of $V$, $E(B-V)$, and the absolute calibration.

\subsubsection{Surface flux} 
\label{sec:sflux}
A number of nLTE model-atmosphere analyses of $\zeta$~Pup have been
published; we use results from \citet{Kudritzki83},
\citet{Bohannan86}, \citeauthor{Puls06} (\citeyear{Puls06},
supplanting \citealt{Repolust04}), \citet{Bouret12}, and
\citet{Pauldrach12}, each of whom employed different, independent
modelling codes.\footnote{The two 20th-century analyses are based on
  plane-parallel, hydrostatic models, without line blanketing;  their
  surface fluxes bracket those of the 21st-century studies, which
  allow for stellar winds and line blanketing.} Each
  study also made different assumptions about the distance to
  $\zeta$~Pup, but we can take the authors' adopted distances and
  derived radii,\footnote{In every case, the authors assumed spherical
    symmetry.} together with the observed flux, to infer their model's
  $V$-band surface flux.  We find
\begin{align*}
F_V (=4\pi H_V) = (6.13 \pm 0.13) \times 10^8\text{~erg cm$^{-2}$ s$^{-1}$
\AA$^{-1}$}.
\end{align*}
Here, as in most of the subsequent analysis, we adopt the standard
deviation of model-atmosphere results as a more conservative, and arguably more credible, estimate
of the true uncertainty associated with the spectroscopic analyses than
is provided by the formal standard error.  In this case the s.d.\  reflects
differences in different modelling procedures,
input physics, and numerical methods, as well as differences in inferred atmospheric parameters
(principally \teff, but also \logg, helium abundance, etc.).

\subsubsection{Effective radius, luminosity} 
\label{sec:diam}
The observed and model
fluxes yield the effective angular diameter directly: $\theta_{\rm eff} = 2\sqrt{f/F} = 0.378 \pm
0.007$~mas, consistent with the observed value of $0.41 \pm
0.03$~mas\footnote{Reduced from the published limb-darkened value of
  0.42 to correct for electron scattering in the wind
  \citep{Kudritzki83}.  We note, however, that formally statistically
  significant discrepancies between the pioneering
  intensity-interferometer results and modern long-baseline optical
  interferometry are not uncommon (cf.\ \citealt{Baines18}).}
(\citealt{HanburyBrown74};
here we use the `effective' qualifier to indicate the result
for a spherical star, of uniform surface flux, that
matches the observed $V$-band brightness of $\zeta$~Pup).

The angular diameter and parallax may be combined to give the effective
radius,
\begin{align*}
\Reff = 13.50 \pm 0.52 \rsun;
\end{align*}
the error
estimate takes into account uncertainties in the parallax, the observed
reddening-free flux, and the surface flux.
Rescaling the luminosities from the nLTE analyses listed in
Section~\ref{sec:sflux} to the \hipp\ distance we find a corresponding
effective luminosity 
\begin{align*}\log(L_{\rm eff}/\lsun) = 5.65 \pm 0.06\end{align*} (where
the quoted error reflects the dispersion in the analyses and the
uncertainty in the distance).

\subsubsection{Preliminary rotation, inclination} 
\label{sec:naive}

For a spherical star the rotation period is 
$P_{\rm rot} = 2\pi \Reff / \veq$, so
an upper limit, $P_{\rm rot}^{\rm max}$, follows from the radius and 
observed \vesini\ by assuming $\sini = 1$.

Observational determinations of \vesini\ are in remarkably good
accord;  the seven values independently determined by
\citet{Kudritzki83}, 
\citet{Bohannan86}, 
\citet{Penny96}, 
\citet{Howarth97}, 
\citet{Repolust04},
\citet{Bouret12}, and 
\citet{Pauldrach12} all lie in the range
203--220~\kms,
averaging $213\pm 7$~\kms\ (s.d).   

[In principle, gravity darkening could result in all
the empirical determinations of \vesini\ systematically
underestimating the true value \citep{Townsend04}, but synthetic
spectra from models such as those discussed in $\S$\ref{sec:prac} indicate that this
effect is negligible for $\zeta$~Pup.]

We therefore have
\begin{align*}
P_{\rm rot}^{\rm max} = 2\pi \Reff / \vesini = 3.21 \pm 0.17\text{ d},
\end{align*}
which securely rules out the 5.1-d rotation period proposed by
\citet{Moffat81}.

If an assumption is made about the rotation period, then a
na\"ive estimate of the axial inclination may  
be obtained instead, from the radius and observed \vesini:
\begin{align*}
\sin(i) = (\prot\, \vesini) / (2\pi\Reff).
\end{align*}
If $\prot = 1.78$~d, then
$\sin(i) = 0.555 \pm 0.028$ ($i=33{\fdg}7 \pm 1{\fdg}9$),
and
$\veq = 384 \pm 15$~\kms,
for the simple, spherical-star case.\newline
[Alternatively, if $\prot = 2.56$~d (the period found in \hipp\
photometry by \citealt{Marchenko98}), then 
$\sin(i) = 0.799 \pm 0.040$ ($i=53{\fdg}1 \pm 3{\fdg}9$),
and
$\veq = 266 \pm 10$~\kms.]

\subsubsection{Spectroscopic mass, pulsation constant}
\label{sec:smass}

Spectroscopic determinations of \logg\ from 
the sources listed in
Section~\ref{sec:sflux} give observed values in the range 3.4--3.6 (dex cgs),
averaging $3.52 \pm 0.08$ (s.d.).  Combining this with the effective
radius established above gives a mass
\begin{align*}
M = 22.1 \pm 4.6\, \msun.
\end{align*}
However, even in the spherical-star approximation, it is possible to
make a statistical correction to the observed surface gravity for the
effects of centrifugal forces,\footnote{\citet{Bouret12} applied this
  correction to obtain their quoted $\logg = 3.64$; we `uncorrected'
  this, using their adopted radius, to infer an observed $\logg =
  3.61$.}  in order to estimate the Newtonian gravity ($GM/R^2$,
sometimes ambiguously referred to as the `true' gravity):
\begin{align*}
g_{\rm N} \simeq g_{\rm obs} + (\vesini)^2/R_*
\end{align*}
\citep{Herrero92, Vacca96, Repolust04}.  For our adopted radius and
\vesini\ (Sections~\ref{sec:diam},~\ref{sec:naive}) this leads to
$\log_{10}(g_{\rm N}) = 3.58$, and a revised mass of
\begin{align*}
M = 25.3 \pm 5.3\, \msun.
\end{align*}
If the photo\-metric period is a pulsation period, 
then the pulsation `constant' is
\begin{align*}
Q & \equiv P_{\rm puls} \sqrt{ \frac{M/\msun}{(R/\rsun)^3} }
  = P_{\rm puls} \sqrt{ \frac{\rsun^2}{G\msun} \frac{g_{\rm
        N}}{\Reff/\rsun} }\\
  & =  0.180 \pm 0.018 \text{ d}
\end{align*}
[or $Q = 0.260\pm0.026$ if $P_{\rm puls} = 2.56$~d].

\subsubsection{Mass-loss rate}

All empirical determinations of the stellar-wind mass-loss rate require 
information on distances; additional uncertainties arise from
ignorance, in detail, of the radial density distribution (e.g., the
acceleration parameter $\beta$ of a canonical velocity law, $v(r) =
v_\infty(1-R_*/r)^\beta$; the degree of clumping, typically
parameterized by a factor $f(r) =
\langle\rho^2(r)\rangle/\langle\rho(r)\rangle^2$; and the overall
geometry).

\citet{Puls06} performed an extensive analysis of multiwavelength
observations (H$\alpha$, IR,
mm, radio), including detailed consideration of clumping; scaling
their results to the \hipp\ distance yields $\mdot = 2.6 \times
10^{-6}$~\msun~yr$^{-1}$ ($\mdot \propto d^{3/2}$).  The result of a
methodologically
independent analysis of X-ray line profiles by \citet{Cohen10}
rescales to $(2.5 \pm 0.2) \times 10^{-6}$ ~\msun~yr$^{-1}$ ($\mdot
\propto d$), in excellent accord.

\subsection{Practicalities:  rapid rotation}
\label{sec:prac}

For a rotation period $\lesssim$3~d we expect rotational effects to be 
non-negligible.
In the Roche approximation (which we adopt here, along with a default
assumption of latitude-independent angular rotation velocity) both the
shape of a star and the ratio of equatorial to polar gravities are
determined solely by \omomc, the ratio of the equatorial angular
velocity to the critical value at which the Newtonian gravitational
force is matched by the centrifugal force, where
\begin{align}
\omcrit = \sqrt{
{(G M)}/ {(1.5 \rpole)^3}
}
\label{eq:vcrit}
\end{align}
for a star of
mass $M$ and polar radius \rpole\
(e.g.,
\citealt{Collins63}).

In the limit that $i = 90^\circ$, then
$\omomc \simeq 0.60$, and the rotational distortions are modest; the
equator is $\sim$2.5~kK cooler than the poles, $\sim$0.12 lower in
\logg, and has a $\sim$7\%\ greater radius (Section~\ref{sec:res1}). However, if the
rotation period really is as short as 1.78~d, then \mbox{$\omomc \gtrsim
0.90$,} and the spherical-star approximation is a rather poor one,
prompting a more thorough treatment.

%
%
%
%

\subsubsection{Model overview}

We model the rotationally distorted, gravity-darkened star by using
\textsc{exoBush} \citep{Howarth01, Howarth16}.  The surface geometry
is that of a Roche equipotential, divided into a large number of
`tiles'.  The specific intensity (or radiance) for each tile is
interpolated from a pre-computed grid of model-atmosphere results, as
a function of wavelength $\lambda$, viewing angle
$\mu$,\footnote{Where $\mu =\cos\theta_{\rm n}$ and $\theta_{\rm n}$ is the angle
  between the surface normal and the line of sight.} local effective
temperature \teffl, and local effective gravity $\loggl$, Doppler
shifted according to the line-of-sight velocity.  Results for all
tiles are summed, weighted by projected area, in order to generate a
synthetic spectrum.  The use of specific intensities means that limb
darkening is taken into account implicitly, in a fully
wavelength-dependent manner.  Gravity darkening is modeled in the
ELR formalism \citep{Espinosa11}, which gives results close to
traditional von~Zeipel gravity darkening \citep{vonZeipel24}, but
which leads to better agreement with, in particular, interferometric
observations (e.g., \citealt{DomdeSou14}).  Intensities are
interpolated in the grid of hydrostatic, line-blanketed, nLTE
\textsc{tlusty} model atmospheres described by \citet{Reeve18}.

\begin{figure}
\centering
\includegraphics[scale=0.45,angle=0]{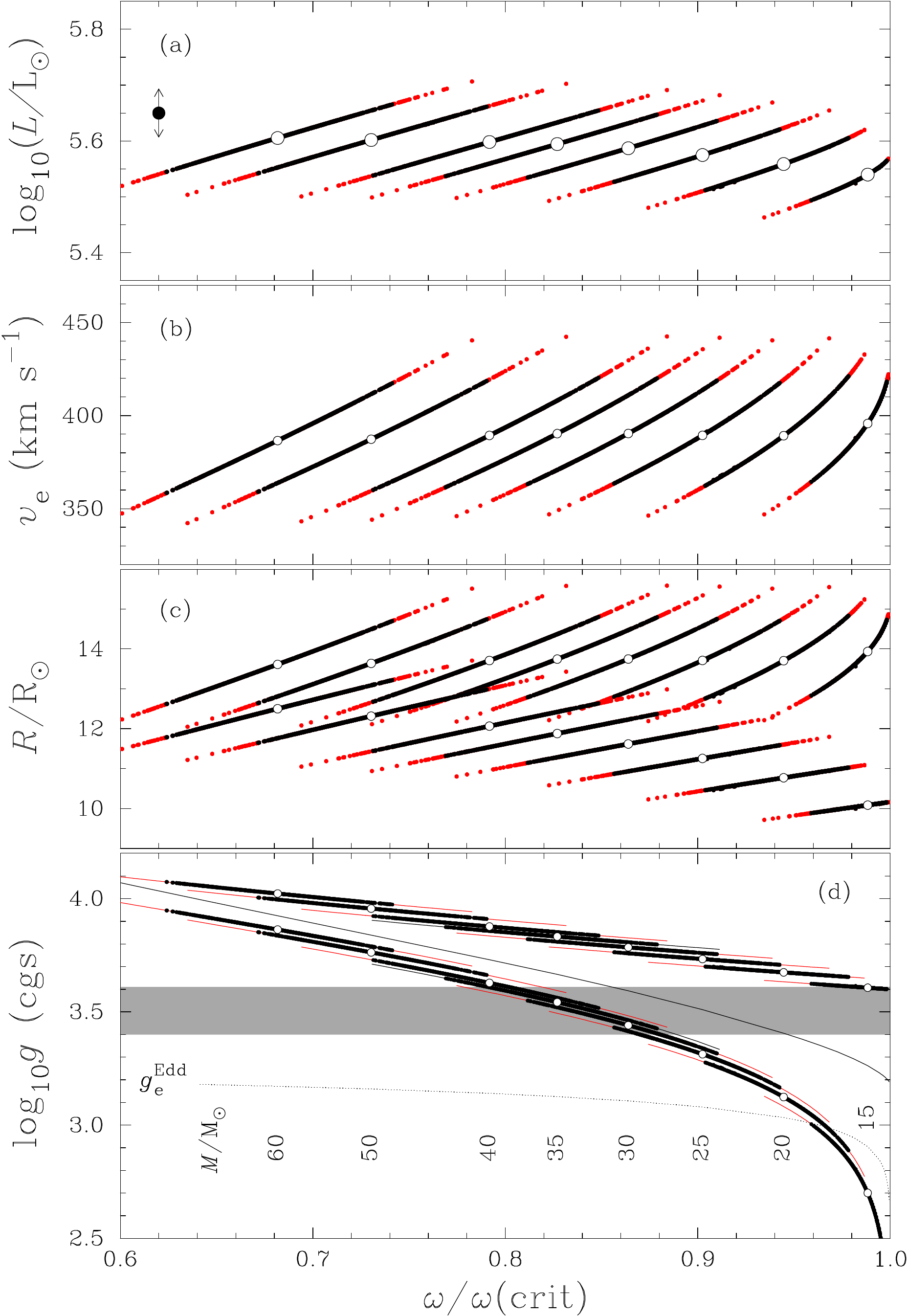}
\caption{Model results as functions of \omomc, assuming a rotation
  period $\prot = 1.78$~d.  For each assumed value of the mass the
  outcomes of 1000 Monte-Carlo replications, generated as described in
  Section~\ref{sec:proc}, are shown as red dots, overlain with the central
  95\%\ of results in black.  Calculations for the adopted central
  values of observed and surface $V$-band fluxes, parallax, and
  \vesini\ are shown as larger open circles, identified by assumed
  mass in panel (d).  \newline 
(a): Luminosity, assuming
  $\teff=40$~kK; the dot and arrows at upper left show the $y$-axis
  displacments resulting from changes of $\pm$1~kK in \teff.  \newline
(b): Equatorial rotation velocity.  \newline 
(c): Equatorial and polar radii (upper, lower sequences), \reqtr, \rpole.  \newline 
(d): Polar and equatorial gravities
  (upper, lower sequences); the loci of mean and
  Eddington gravities, defined in Section~\ref{sec:res1}, are shown as
  continuous and dashed lines, respectively.  The horizontal grey band
  indicates the full range of spectroscopically determinded surface
  gravities from the analyses listed in Section~\ref{sec:sflux}.  }
\label{fig:BigPic}
\end{figure}

\subsubsection{Procedure}
\label{sec:proc}

The calculations essentially follow the steps outlined in
Section~\ref{sec:prelim}, but make full allowance for the rotational
distortion and gravity darkening, using a Monte-Carlo (MC) approach to
observational uncertainties.  

For each MC realisation we first generate a set of values for the observed flux,
surface flux, and parallax (each drawn from the corresponding mean
values and errors given previously, assuming Gaussian distributions)
in order to establish a value for the effective radius.

We combine this radius with a model-atmosphere \mbox{$\sim$$V$-band}
surface flux\footnote{We actually use the monochromatic flux at
  546.5~nm; the exact choice is of no consequence as long as a
  line-free wavelength is chosen, as is the case here.} to generate a
pseudo-`observed' target flux (cf.\ Section~\ref{sec:aside}).   A
\vesini\ value is also drawn from the adopted distribution.

Further steps proceed according to two alternative assumptions about
the star's rotation:

\goodbreak
\newcounter{itemcounter}
\begin{list}
{\textit{(\roman{itemcounter})}}
{\usecounter{itemcounter}\leftmargin=0em}
\item{\textit{Assuming $i = 90^\circ$.}}\\
If $\zeta$~Pup is viewed close to equator-on (as has frequently been
assumed, based on its exceptionally large \vesini\
value), then rotational distortions are sufficiently small that it
may be considered reasonable to infer a mass from the observed gravity,
\begin{align*}
M \simeq g_{\rm N} \rpole\reqtr/G
\end{align*}
(where a
specific `observed' \loggN\ value is generated for each MC cycle, and
\reqtr\ is the equatorial radius).
We compute a full, rotationally-distorted, gravity-darkening
model for this mass (and $i=90^\circ$), taking $\rpole \simeq \Reff$
(in practice,
slightly less) as a first estimate.  The flux from this model will not
match the target flux, because of the rotational effects now
included; so we rescale the radii appropriately, recalculate the mass,
and compute a new model. The process is then iterated until the model
flux matches the target flux,
resulting in a self-consistent pair of $\rpole, M$ values that
reproduce the target flux and \vesini, for the assumed inclination.\newline
\item{\textit{Assuming $\prot = 1.78$~d.}}

In this case, several sequences of models are run, each
characterized by a specified, assumed mass (because `the' surface
gravity is a poorly defined quantity for the implied rapid rotation).  For each
sequence, the rotation period gives an initial estimate of the
inclination.
As before, we generate full models, but now iterate to identify
a self-consistent pair of $\rpole, i$ values that
reproduce the target flux and \vesini, for the assumed mass.

\end{list}

\begin{table*}
  \caption{Selected model results.  The second column
    summarizes the discussion of Section~\ref{sec:prelim}.
Subsequent columns
    report the more through analyses described in
    Section~\ref{sec:prac}, under the assumptions stated in the
    column headers. The equatorial
    Eddington gravity, $g^{\rm Edd}_{\rm e}$, is defined in Section~\ref{sec:res1}.}
\begin{center}
\begin{tabular}{l    r @{\,}c@{\,} l    r @{\,}c@{\,} l  r @{\,}c@{\,} l  r @{\,}c@{\,} l  r @{\,}c@{\,} l }
\hline
& \multicolumn{3}{c}{Spherical \T\B} & \multicolumn{3}{c}{$i
  = 90^\circ$} &
\multicolumn{9}{c}{\quad\leaders\hbox{\rule[0.2em]{.1pt}{0.4pt}}\hfill\mbox{}\;
  $\prot
  = 1.78$~d \;\leaders\hbox{\rule[0.2em]{.1pt}{0.4pt}}\hfill\mbox{} \quad} \\
\hline
$M/\msun$\T     & 25.3 & $\pm$ & 5.3     & 26.6  &$\pm$ & 5.6 & \multicolumn{3}{c}{$\equiv 15$} & \multicolumn{3}{c}{$\equiv 25$} & \multicolumn{3}{c}{$\equiv 50$} \\
\rpole/\rsun  & \multirow{2}{*}{13.50} & \multirow{2}{*}{$\pm$} & \multirow{2}{*}{0.52} 
                                       & 13.22 &$\pm$ & 0.54 & 10.06 &$\pm$ & 0.07 & 11.25 &$\pm$ & 0.19 & 12.32 &$\pm$ & 0.34 \\
\reqtr/\rsun  & \multicolumn{3}{c}{}   & 14.08 &$\pm$ & 0.53 & 13.86 &$\pm$ & 0.05 & 13.72 &$\pm$ & 0.49 & 13.66 &$\pm$ & 0.54 \\
$T_{\rm p}/$~kK$^\ddagger$
              & \multicolumn{3}{c}{40}             & 41.59& $\pm$ & 0.32  & 46.21  &$\pm$ & 0.38 & 44.35  &$\pm$ & 0.34 & 42.47  &$\pm$ & 0.25 \\
$T_{\rm p}/T_{\rm e}$
              & \multicolumn{3}{c}{$\equiv 1.0$}   & 1.064&$\pm$ & 0.014  & 1.392  &$\pm$ & 0.064 & 1.207 &$\pm$ & 0.023 & 1.103  &$\pm$ & 0.012 \\
\rule{0pt}{3ex}$\log_{10}\left({L/\lsun}\right)^\ddagger$
              & 5.65 & $\pm$ &0.06     & 5.641  &$\pm$ &0.033 & 5.536  &$\pm$ &0.019 & 5.575  &$\pm$ &0.024 & 5.602  &$\pm$ &0.030 \\
\loggp, cgs   & \multirow{2}{*}{ 3.58} & \multirow{2}{*}{$\pm$} & \multirow{2}{*}{0.08} 
                                       & 3.611  &$\pm$ & 0.078  & 3.609  & $\pm$ & 0.006 & 3.734  &$\pm$ & 0.014 & 3.956  &$\pm$ & 0.024 \\
\logge, cgs   & \multicolumn{3}{c}{}   & 3.494  &$\pm$ & 0.103  & 2.695  & $\pm$ & 0.189 & 3.309  &$\pm$ & 0.069 & 3.761  &$\pm$ & 0.048 \\

$\log_{10}(g^{\rm Edd}_{\rm e})$, cgs
              & \multicolumn{3}{c}{3.23}
              & \multicolumn{3}{c}{3.19}
& 2.905  &$\pm$ & 0.065 & 3.080  &$\pm$ & 0.019 & 3.159  &$\pm$ & 0.008 \\
\rule{0pt}{3ex}\omomc\       & \multicolumn{3}{c}{---}                        
                                      & 0.602 &$\pm$ & 0.053 & 0.985  &$\pm$ & 0.010 & 0.902  &$\pm$ & 0.022 & 0.731  &$\pm$ & 0.030 \\
Inclination $i\;(^\circ)$   
              & \multicolumn{3}{c}{---}                        
                                      & \multicolumn{3}{c}{$\equiv 90$} & 32.8  &$\pm$ & 1.7 & 33.2  &$\pm$ & 1.8 & 33.4  &$\pm$ & 1.9 \\
\veq\ (\kms)  & 213 & $\pm$ & 7 & 213 & $\pm$ & 7 & 394 &$\pm$& 14 & 390 &$\pm$& 14 & 388 &$\pm$& 15 \\

\prot\ (d) \B & $\leq3.21$ & $\pm$ & 0.17 
                                      & 3.35  &$\pm$  &0.16 &
                                      \multicolumn{9}{c}{\quad \leaders\hbox{\rule[0.2em]{.1pt}{0.4pt}}\hfill\mbox{}
                                        $\;\equiv 1.78\;$ \leaders\hbox{\rule[0.2em]{.1pt}{0.4pt}}\hfill\mbox{} \quad}  \\
\hline
\multicolumn{16}{l}{
\begin{minipage}[t]{1.44\columnwidth}
{\rule{0pt}{3ex}$^\ddagger$Assuming $\teff = 40$~kK, excepting
  the spherical-star luminosity; if $f = \teff^*/40$~kK, where $\teff^*$
  is the true effective temperature, then $T^*_{\rm p} = fT_{\rm p}$
  and $\log_{10}\left({L^*/\lsun}\right) =
  \log_{10}\left({L/\lsun}\right) + 4 \log_{10}f$.
Radii at $\prot =1.78$~d are smaller than for the $i=90^\circ$ models for
the reason mentioned in Section~\ref{sec:aside} -- the hotter polar
regions are more clearly presented to the observer at $i \simeq 33$
than equator-on, so a smaller emitting area is required to match the
observed flux, for given \teff.} 
\end{minipage}
}
%
\end{tabular}
\end{center}
\label{t_obs}
\end{table*}

\subsection{An ignorable aside on model-atmosphere intensities and fluxes}
\label{sec:aside}

In principle, the `target flux' used in Section~\eqref{sec:proc} could be matched to
the adopted value for the observed flux, simply by adjusting the
model's (global) effective temperature, defined as
\begin{align*}
\teff = \sqrt[4]{{\int{\sigma(\teffl)^4\,\text{d}A}}\left/{{{\int{\sigma\,\text{d}A}}}}\right.}
\end{align*}
(where $\sigma$ is the Stefan--Boltzmann constant and the integrations
are over total surface area).  However, this adjustment is
unnecessary as long as only the factors involving geometry are of
interest (i.e., mass, radius, inclination, \omomc); the requirement
then is only that a \textit{consistent} temperature be adopted, not
that it be `correct' (beyond first order).  We simply adopt
$\teff = 40$~kK, representative of results from the detailed analyses
mentioned in Section~\ref{sec:sflux} (even though our \textsc{tlusty}
model fluxes will not precisely match the fluxes of the models used
in those analyses).

Nevertheless, there remains a minor inconsistency in our modelling of
the \mbox{1.78-d} rotation constraint,
which arises because the relationship between the \textit{perceived}
temperature and the global effective temperature varies with
inclination -- a gravity-darkened star will generally appear hotter if
viewed pole-on than equator-on.  Consequently, as the inclination
changes from one iteration to the next, the perceived temperature
changes, at constant \teff\ (as does the computed observed flux, even
at constant radius).

This could be corrected for, given an appropriate prescription for
transforming between effective and perceived temperature, but one
would first need to define the latter quantity (e.g., by synthesizing
the full gravity-darkened spectrum and then modelling it as though it
arose from a spherical star of uniform surface intensity; the result
would still depend on the analysis criteria).  In practice, for the
rather small range of inclinations that our models are found to span,
the $V$-band flux variation from this effect is negligible (acting as
additional source of very-low-amplitude noise in the radius determinations).

\subsection{Results}
\label{sec:res1}

Some results of the models are summarized in Table~\ref{t_obs}.  The
$i= 90^\circ$ models represent one geometrical extreme, and provide
secure upper-limit values for \rpole, \reqtr, and \prot, along with
lower limits to \veq\ and \omomc.  The 2-$\sigma$ upper limit on the
rotation period is $\prot <3.7$~d, ruling out previous suggestions that
the value may be $\sim$5.1~d \citep{Moffat81}.

Additional results for $\prot=1.78$-d models are shown in
Fig.~\ref{fig:BigPic}.  These models provide some limits on the 
allowed masses for this rotation period.  First, 
$M \gtrsim 12.8\msun$ is required for $\omomc \le 1$.  Secondly, the `Eddington gravity' required to retain
material against the radiation force is
\begin{align*}
g_{\rm Edd} &= \frac{\sigma T^4}{c}\kappa\\
            & \simeq 6.56 \times 10^{-16}\, (T/\text{K})^4\text{ cm s}^{-2}
\end{align*}
where $\kappa$ is the flux-mean opacity (per unit mass) and 
$c$ is the speed of light;  the numerical value
equates $\kappa$ with the electron-scattering opacity for
a fully-ionized solar-abundance mix.    The equatorial value,
$g_{\rm e}^{\rm Edd}$, is plotted in
Fig.~\ref{fig:BigPic};  it is exceeded by the actual equatorial gravity
only for $M \gtrsim 18\msun$.\footnote{We are aware that this has implications for the
topology of the equipotential surface that are neglected in the
present work.   However, for traditional \citeauthor{vonZeipel24} gravity darkening, there is
no effect \citep{Howarth97o}, and it is only at near-critical rotation
that the ELR formalism departs significantly from this.  If anything,
radiation-pressure effects will render our lower limits to mass more secure.}

Also shown in Fig.~\ref{fig:BigPic} is a mean gravity, 
$\overline{g} = \int{\gloc\,\text{d}a} / \int{\text{d}a}$, where 
\gloc\ is the local gravity and the
integration is over the projected area.
If we suppose that the observed \logg\ values fall between
the models' equatorial and polar values then $M \lesssim 40\msun$;  or if we
speculate that the observed gravity can be identified
with the mean gravity, then $20\lesssim M/\msun \lesssim 30$.

Table~\ref{t_obs} includes detailed results for models at 15, 25, and
50\msun, representing what we consider to be the very extreme range of plausible masses
if $\prot=1.78$~d, together with a `best guess' central value.  These
models are intended to illustrate the sensitivity, or otherwise, of various parameters
to the assumed mass.\footnote{In principle, the synthetic spectra
  generated as part of the modelling could be used to
  gravity-sensitive lines, such as the wings of H$\gamma$, to better
  constrain the mass.  Unfortunately, it isn't possible to achieve
  convergence for line-blanketed, hydrostatic models for gravities
  less than $\logg \lesssim 3.5$ at the relevant temperatures; in our
  \textsc{exoBush} modelling, where necessary we used the lowest
  available gravity.}  The inferred inclination and equatorial
velocity are particularly robust, with mean values averaged over \textit{all}
models, 15--50\msun, being $i = 33{\fdg}2 \pm 1{\fdg}8$, $\veq = 390
\pm 16$~\kms.

The implied axial inclination if $\prot=1.78$~d is unremarkable, but
the equatorial velocity would be unprecedented for an early O$\;$I
star; in the Galaxy, only a handful of near-main-sequence, late-O stars have
comparable (projected) rotation velocities.\footnote{ HDs~93521,
  149757, and the ONn stars 
\citep{Howarth01, Walborn11, Martins15}}.

\begin{figure}
\centering
\includegraphics[scale=0.43,angle=270]{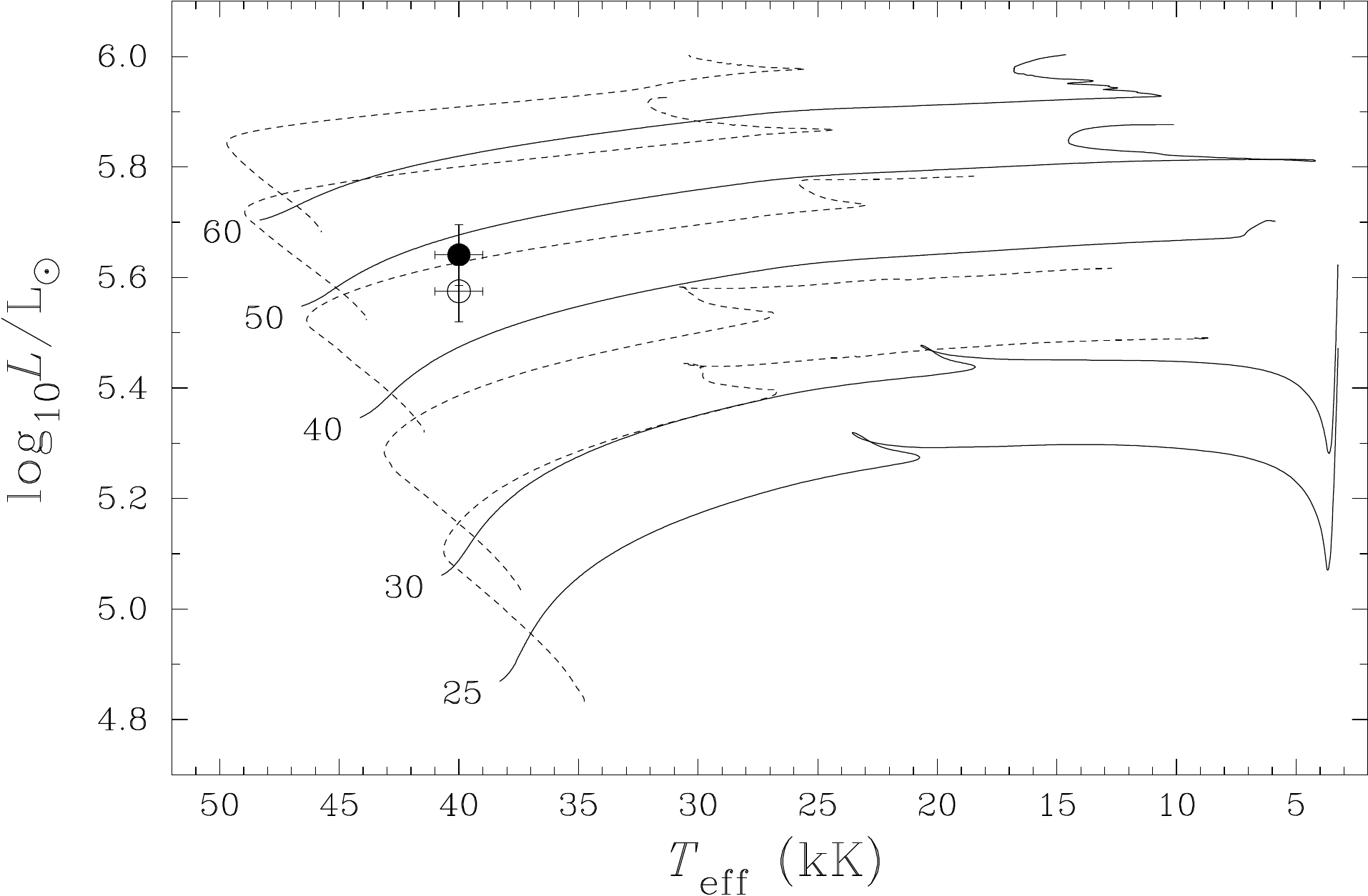}
\caption{Evolutionary tracks from \citet{Brott11}, labelled by ZAMS
  mass in solar units.  Continuous and dotted lines are for models
  without rotation and with initial equatorial rotational velocities of
  $\sim$550~\kms.  Filled, open circles: the location of $\zeta$~Pup
  for $i=90^\circ$ and for $\prot = 1.78$~d, $M=25\msun$ respectively
  (cf.\ Table~\ref{t_obs}), at $\teff = 40$~kK; the `error
  bars' indicate a range of $\pm$1~kK in \teff.}
\label{fig:HRD}
\end{figure}

\begin{table}
  \caption{\textsc{Bonnsai} model results (cf.\ Section~\ref{sec:disco}).
Small open circles indicate parameters used as inputs (with
observational values and 1-$\sigma$ gaussian errors listed in
column~2);
quantities assessed on the ZAMS are indicated with a subscript `1'.}
\begin{center}
\begin{tabular}{l   r@{\,}c@{\,}l  r@{\,}r@{\,}l  r@{\,}r@{\,}l r@{\,}r@{\,}l }
\hline
\multicolumn{1}{c}{Parameter\T\B}&\multicolumn{3}{c}{Input}
&\multicolumn{6}{c}{\quad\leaders\hbox{\rule[0.2em]{.1pt}{0.4pt}}\hfill\mbox{}\;\;Replicated Observables\T}\leaders\hbox{\rule[0.2em]{.1pt}{0.4pt}}\hfill\mbox{}\quad\;\;\\
&&
&&\multicolumn{3}{c}{\;Model 1}&\multicolumn{3}{c}{\;Model 2}\\
\hline
\teff/kK              &40&$\pm$&1         &$\circ$& $39.92$&$^{+1.16}_{-0.88}$ &$\circ$& $40.08$&$^{+0.95\T}_{-1.11}$ \\
\logL     \T          &5.60&$\pm$&0.05    &$\circ$& $5.58$&$^{+0.05}_{-0.05}$  &$\circ$& $5.58$&$^{+0.06}_{-0.04}$    \\
\vesini/\kms \T       &213&$\pm$&7        &$\circ$& $210$&$^{+12}_{-6}$        &$\circ$& $210$&$^{+13}_{-5}$          \\
$Y$          \T       &0.41&$\pm$&0.05    &       & $0.27$&$^{+0.04}_{-0.01}$  &$\circ$& $0.39$&$^{+0.09}_{-0.06}$    \\
\veq/\kms    \T       &    &     &        &       & $220$&$^{+58}_{-15}$       &       & $270$&$^{+36}_{-17}$         \\
$v_{\rm e, 1}$/\kms \T&&&                 &       & $260$&$^{+114}_{-29}$      &       & $440$&$^{+47}_{-40}$         \\
$M_{\rm evol}/\msun$ \T        &&&          &       & $42.4$&$^{+3.4}_{-3.6}$    &       & $37.0$&$^{+3.8}_{-2.9}$    \\
$M_1/\msun$     \T      &&&                 &       & $45.4$&$^{+3.1}_{-3.6}$    &       & $41.8$&$^{+3.6}_{-2.7}$    \\
Age/Myr         \T\B    &&&                 &       & $2.20$&$^{+0.57}_{-0.49}$  &       & $3.56$&$^{+0.77}_{-0.75}$  \\
\hline
\end{tabular}
\end{center}
\label{t_bonn}
\end{table}

\subsection{Evolutionary status}
\label{sec:disco}

The \hipp\ distance constrains the radius, hence luminosity,
reasonably tightly, allowing us to locate $\zeta$~Pup quite precisely
in the H--R diagram (Fig.~\ref{fig:HRD}), independently of its
supergiant
spectral classification.  It is evident, simply by
inspection, that our empirically inferred preferred mass range,
$\sim$20--30\msun, is inconsistent with the evolutionary mass implied
by the \citet{Brott11} models, $M_{\rm evol} \sim 40$--50\msun\
(rapidly-rotating$\rightarrow$non-rotating progenitor).

We can elaborate this inference with a
\textsc{Bonnsai} analysis (\citealt{schneider14}\footnote{The BONNSAI
  web service is available at\newline
  \url{http://www.astro.uni-bonn.de/stars/bonnsai}}), built on the same
evolutionary models.  For a minimal set of
observational constraints
($\teff = 40 \pm 1$~kK, $\logL =  5.60 \pm 0.05$, $\vesini = 213 \pm
7$~\kms, and a default
set of priors), the single-star evolutionary tracks imply
$M_{\rm evol}/\msun = 42.4^{+3.4}_{-3.6}$ (Table~\ref{t_bonn}, `Model~1').

This minimal model has a predicted surface-helium abundance that is
essentially solar, in conflict with observations; the sources listed
in Section~\ref{sec:sflux} give helium abundances by number in the
range $y = 0.14$--0.20, averaging $0.17 \pm 0.02$ (s.d.; i.e., mass
fraction $Y \simeq 0.41 \pm 0.05$).  A solution is still possible
after imposing this additional constraint (`Model~2'), with $M_{\rm evol}/\msun =
37.0^{+3.8}_{-2.9}$.

However, imposing the  further constraint of $\veq = 390
\pm 16$~\kms\ allows for no acceptable \textsc{Bonnsai} solutions (at any $Y$).

The discrepancies between empirical and single-star evolutionary masses
support the proposal by \citet{vanRensbergen96} that the runaway
and rotational properties of $\zeta$~Pup are most readily understood
in the context of previous binary interaction, and not single-star
evolution. Although recent versions of this `Brussels scenario' target
high-mass solutions for $\zeta$~Pup, with $M \gtrsim 60$\msun (e.g.,
\citealt{Vanbeveren12}, \citealt{Pauldrach12, TahinaR18}), alternative 
channels can lead to undermassive (or overluminous) runaways
\citep{Vanbeveren94}.

\section{Starspots?}
\label{sec:Spots}

Non-radial pulsations and corotating starspots are the two most
obvious candidates for the processes underpinning the 1.78-d
photo\-metric variability observed
in $\zeta$~Pup
(cf.\ $\S$3.4.1 of
\citealt{TahinaR18} for a thorough discussion).
\citet{Howarth14} suggested that the physical origin of
the signal may be pulsation associated with low-$\ell$ oscillatory
convection modes, noting rough consistency with theoretical models by
\citet{Saio11}.  Our revised physical parameters render the
comparison problematic; \citeauthor{Saio11}'s stability analysis was
based on structures computed for standard single-star evolutionary
tracks, which don't explore the parameter space that now appears
pertinent to $\zeta$~Pup ($\teff \simeq 40$~kK, $M \simeq 25 \msun$,
with potentially strong mixing in the scenario sketched by \citealt{Vanbeveren94}).

By contrast, \citet{TahinaR18} advocated a rotationally-modulated `hotspot'
interpretation of the \mbox{1.78-d} period, 
and on that basis used light-curve inversion
techniques to map the required surface-brightness distribution.  Our
updated understanding of the stellar geometry allows us to re-examine
this question, and in particular to assess if line-profile variability
offers a potential test of the rotational-modulation hypothesis.

\subsection{Photometry}
\label{sec:sphot}

\begin{figure}
\centering
\includegraphics[scale=0.42,angle=270]{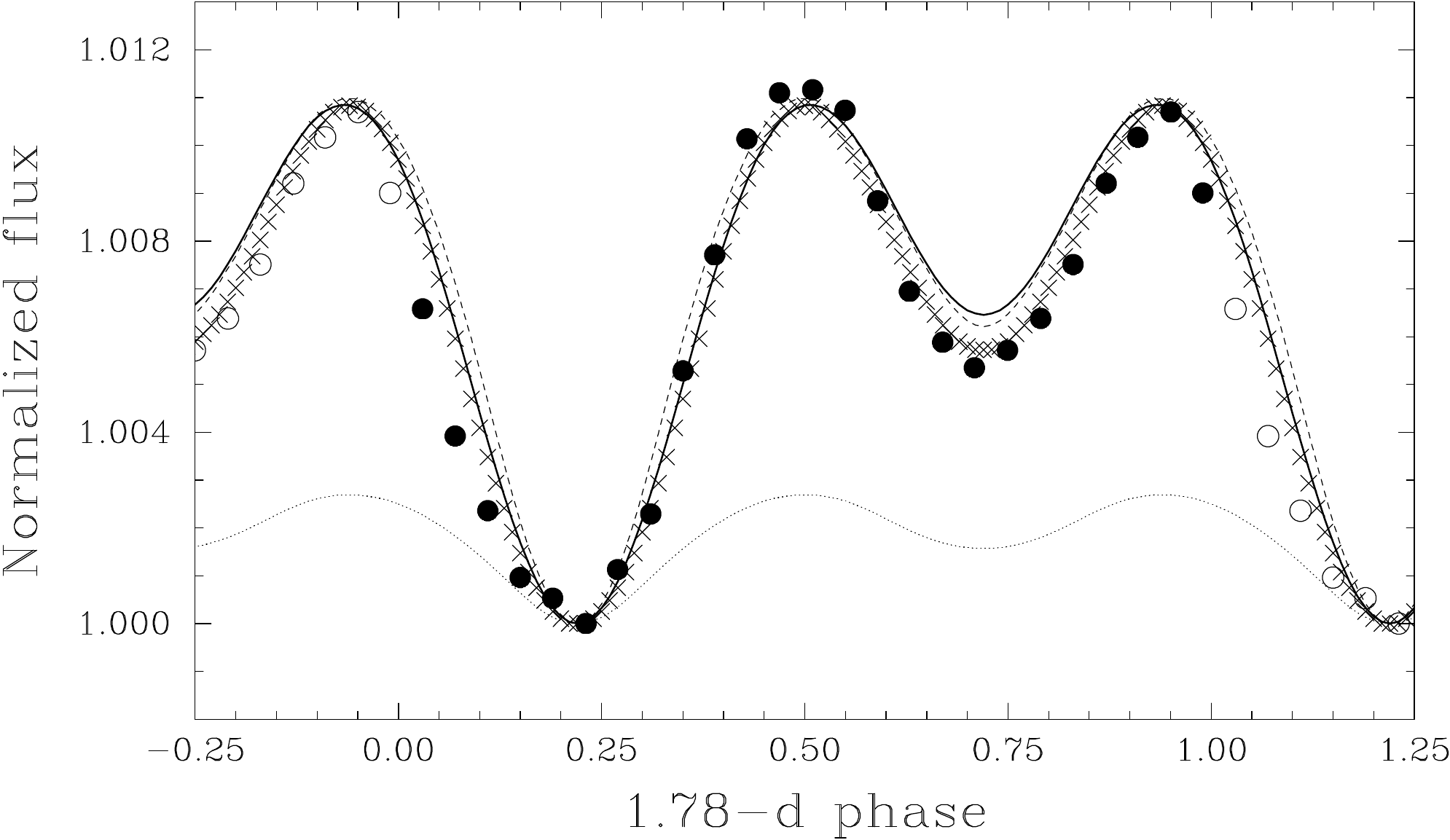}
\caption{\textit{BRITE-Constellation} photometry (filled \& open
  circles, from \citealt{TahinaR18}) compared to simple two-starspot
  models.  Phase 0.5 is arbitrarily chosen
  to correspond to the leading (hot) spot transiting the meridian, and
  fluxes are normalized to a minimum value of unity. 
Values of ($\alpha_{\rm s}/^\circ$, $T_{\rm s}$/kK) for hotspot models
are:  dotted
  (low-amplitude) line, (10, 42.5);  dashed line (26, 42.5);
solid line (14, 50.0).  The cool-spot model discussed in
$\S$\ref{sec:sphot} is shown as small `$\times$' symbols.}
\label{fig:Spot}
\end{figure}

For an initial exploration, we took a representative model with
$\prot=1.78$~d and $M=25\msun$ (cf.\ Table~\ref{t_obs}).  We modified
\textsc{exoBush} to include starspots that subtend a constant angular
radius $\alpha_{\rm s}$ at the centre of mass (i.e., are approximately
circular on the surface), at constant temperature $T_{\rm s}$.  For
specificity, we compare results with \citeauthor{TahinaR18}'s
`Part~IV' photo\-metric dataset, which has a moderately, but not
exceptionally, large amplitude ($\sim$1\%; Fig.~\ref{fig:Spot}) and
their corresponding $i=33^\circ$ surface-brightness reconstruction (cf.\ panel~2 of their
Fig.~10).  This reconstruction has a rather simple hotspot geometry; we
approximate their results with two $\sim$equatorial spots (colatitudes
$\theta_{\rm s} = 85^\circ$), separated by $158^\circ$ in longitude.  While this
spot model is not intended as a detailed `best fit', it
captures the essential characteristics of the observations at this
epoch.

Based on the \citet{TahinaR18} reconstruction, values of $\alpha_{\rm
  S} = 10^\circ$, $T_{\rm s} = 42.5$~kK were first employed; the
results of the ($V$-band) photo\-metric predictions are confronted with
observations in Fig.~\ref{fig:Spot}.  They substantially underestimate
the observed amplitude; we believe this to be a straightforward
consequence of the fact that the inversion approach is essentially a
mathematical methodology intended simply to recover a
surface-brightness distribution, while ours is a direct physical
model.

To recover the amplitude observed in the \textit{BRITE-Constellation}
photometry requires spots that are either significantly
larger  or
significantly hotter than initially assumed.  
We find that
$\alpha_{\rm s} \simeq 26^\circ$ for $T_{\rm s} = 42.5$~kK;
$T_{\rm s} = 50$~kK requires $\alpha_{\rm s} \simeq
14^\circ$.  

Although            these models reproduce the general characteristics of the
photometry, we recall that it is always possible to
construct a spot model capable of reproducing periodic, low-amplitude
photometric
variability; and a successful model fit (or inversion) based on the
assumption of surface hotspots is not a proof of their existence, but
is only a plausibility check.  To emphasize this point, we have
constructed a simple illustrative model with two \textit{cool} spots,
$T_{\rm s} = 30$~kK, $\alpha_{\rm s}(1, 2) = (17, 20)^\circ$, separated by
180$^\circ$ in longitude.   This \textit{ad hoc} model matches the data at least as
well as the hotspot calculations (Fig.~\ref{fig:Spot}).

\subsection{Line-profile variability}
\label{sec:lpv}

\begin{figure}
\centering
\includegraphics[scale=0.345,angle=270]{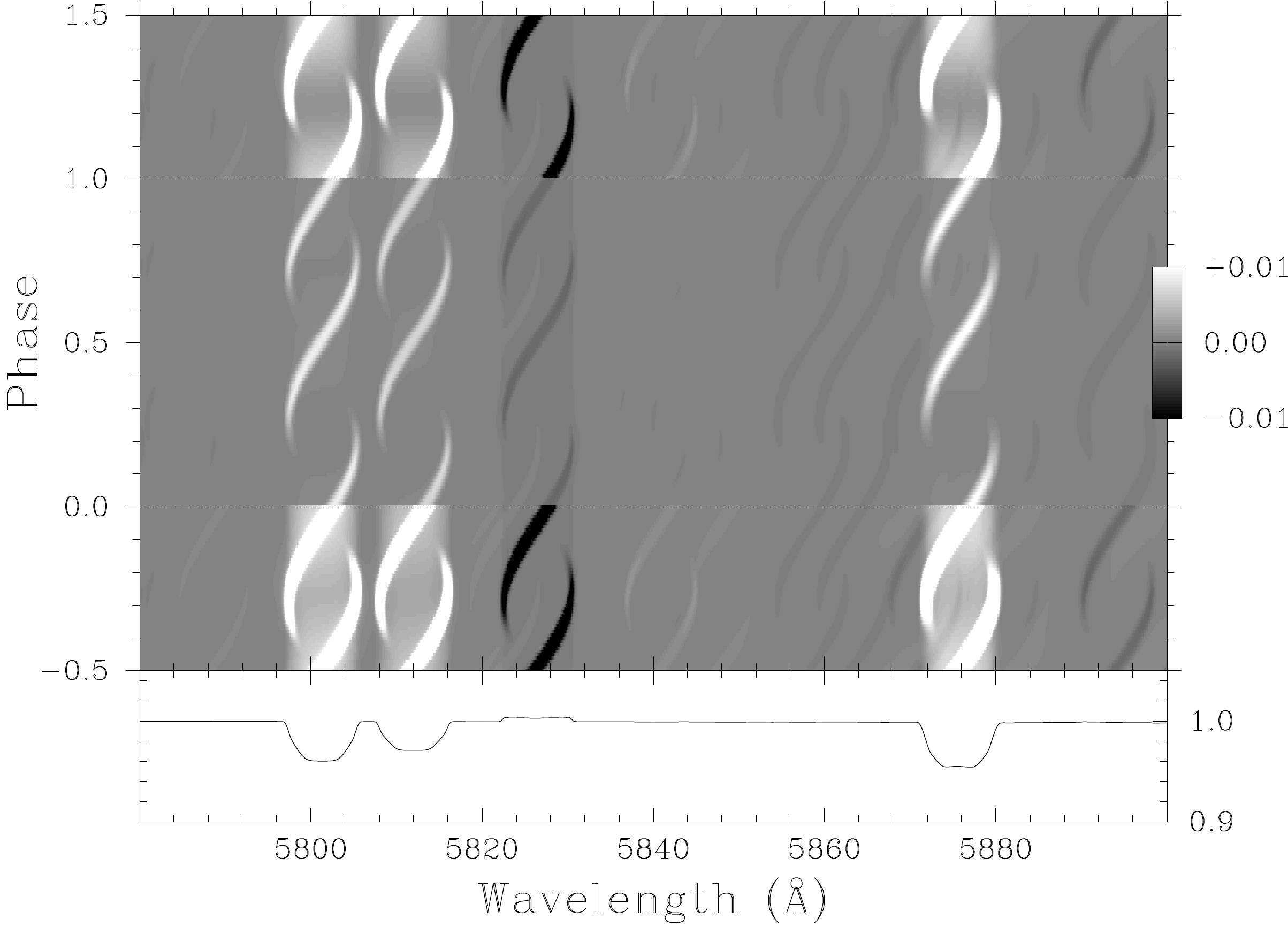}
\caption{Dynamic spectrum of predicted line-profile variability
($\S$\ref{sec:lpv}),
  showing differences in rectified spectra from the unperturbed state.
The embedded intensity scale calibrates the central subpanel;  the
upper and lower
subpanels are `stretched' by a factor 10 in intensity in order
to emphasize low-amplitude features.
The
  strongest modulations arise from C\,\textsc{iv}~$\lambda\lambda$5801,~5812; 
C\,\textsc{iii}~$\lambda$5826; and
He\,\textsc{i}~$\lambda$5876.\newline
Lower panel:  rectified unperturbed spectrum.}
\label{fig:gscale}
\end{figure}

A potentially more stringent test of the nature (and existence) of
any starspots is offered by the spectroscopic line-profile variability
that they should generate.    An exploratory `proof of concept'
calculation is shown in Fig.~\ref{fig:gscale}, based on the 
$2\times (T_{\rm s} = 50$~kK, $\alpha_{\rm s} =
14^\circ)$ spot model.

Cool spots generally give rise to (pseudo-)emission `bumps' in
spectrally resolved line profiles, essentially because they remove
less flux at the projected spot velocity than in the continuum (e.g.,
\citealt{Vogt83}).  Bright spots will give rise to `absorption dips'
by the same reasoning, but only as long as the line equivalent
widths do not change by a significant amount between the spot and
adjacent unperturbed photosphere.

That is not the case for the model presented here.  The assumed
temperature contrast, $T_{\rm e}$\,:\,$T_{\rm s} \sim 37$:50, is
strong enough to result in substantial changes in line strength.  For
example, C\,\textsc{iii}~$\lambda$5826 is a weak emission line in the
intrinsic equatorial spectrum, but disappears entirely in the
model's high-temperature spot; similarly, He\,\textsc{i}~$\lambda$5876
absorption weakens greatly.  Moreover, the spots are hot, at low
gravity, and viewed relatively far from normal incidence.  In the
hydrostatic, line-blanketed models used here, these circumstances are often
accompanied by substantial decreases in the strengths of absorption
lines, which may even go into emission through nLTE effects, as occurs for
C\,\textsc{iv}~$\lambda\lambda$5801,~5812.  Consequently, when
rectified and differenced with the unperturbed spectrum, the signature
variability of hot spots in the models can be in the form of either
absorption or emission features (Fig.~\ref{fig:gscale}).

Whether or not these modelled line-strength characteristics are
quantitatively reliable, there is considerable diagnostic potential in
the straightforward dynamical content of velocity- and
temporally-resolved spectroscopy.  Most importantly,
Fig.~\ref{fig:gscale} illustrates a specific discriminant between
traditional, $i \simeq 90^\circ$, models and the low-inclination,
$\prot\ = \pphot$
alternative.  For $\sim$equator-on configurations, longitudes more
than $90^\circ$ from the central meridian are never visible, but at
lower inclinations (as required by rotational modulation), features
that occur 
in the hemisphere
nearer the observer
(i.e., `north' of the equator)
are visible
beyond $\pm$0.25 in phase either side of transit, at submaximal
velocity excursions, generating `\sqigl'-shaped features 
with
red-to-blue `tails'  
in the
dynamic spectrum.

This is true even for the near-equatorial ($\theta_{\rm s}=85^\circ$),
slightly extended spots used in the exploratory model.  Spots with smaller
colatitudes would be visible for greater fractions of the
rotation period (giving relatively bigger tails to the \sqigl\ tracks), and
would also leave their signature over smaller ranges in velocity
($\Delta{v_{\rm s}}/\vesini \sim \sin\theta_{\rm s}$).

The predicted amplitudes of the spectroscopic spot signatures are on the
order of $\sim$1\%, which should be comfortably observable.
\citeauthor{TahinaR18} (\citeyear{TahinaR18}; their $\S$4.1.3) report
recovering the \mbox{1.78-d} signal in photospheric absorption lines,
including C\,\textsc{iv}~$\lambda\lambda$5801,~5812, but with ``no
obvious pattern'', which may be seen as weak evidence against rotational
modulation.  Other published time series also show only marginal evidence,
at best, for any \mbox{$\sim$1.8-d} periodicity \citep{Reid96,
  Berghofer96}, although we have no way of knowing if the photometric
signal was present at the time of those observations (cf.\
$\S$\ref{hipp:phot}).

\section{Photospheric and stellar-wind variability}

\subsection{Corotating Interaction Regions and Discrete Absorption Components}
\label{sec:cirdac}

Ultraviolet spectroscopy with the International Ultraviolet Explorer
(IUE; \citealt{Boggess78a})
showed that `discrete
absorption components', or DACs, are a ubiquitous characteristic of
early-type stars with strong winds.  They are characterized by
red-to-blue migration of features through the absorption components of
P-Cygni profiles, with accelerations that are significantly slower than
expected for the ambient outflow (e.g., \citealt{Prinja86, Prinja88,
  Kaper96, Kaper99}).

The slow acceleration, in particular, has prompted an interpretation
in terms of corotating interaction regions (CIRs; \citealt{Mullan84,
  Mullan86, Cranmer96}).  This interpretation is bolstered by the
(rather loose) anticorrelation between DAC recurrence timescales and
\vesini\ values for individual stars (\citealt{Prinja88a, Kaper99,
  Howarth07}).  However, in most cases the DACs are not strictly
periodic,\footnote{An apparently periodic feature has been observed in
  UV \mbox{P-Cygni} profiles of the rapidly rotating B0.5\,Ib star
  HD~64760, but this seems to be distinct from classical DACs
  \citep{Prinja95, Fullerton97}.} so it remains open to debate as to
whether the DACs are initiated directly in corotating (or very nearly
corotating) photospheric features, or develop as an intrinsic property
of radiatively-driven winds (cf., e.g., \citealt{Martins15a}),
notwithstanding that DAC-like behaviour has been traced to rather low
velocities \citep{Massa15}.

\begin{figure}
\centering
\includegraphics[scale=0.4,angle=0]{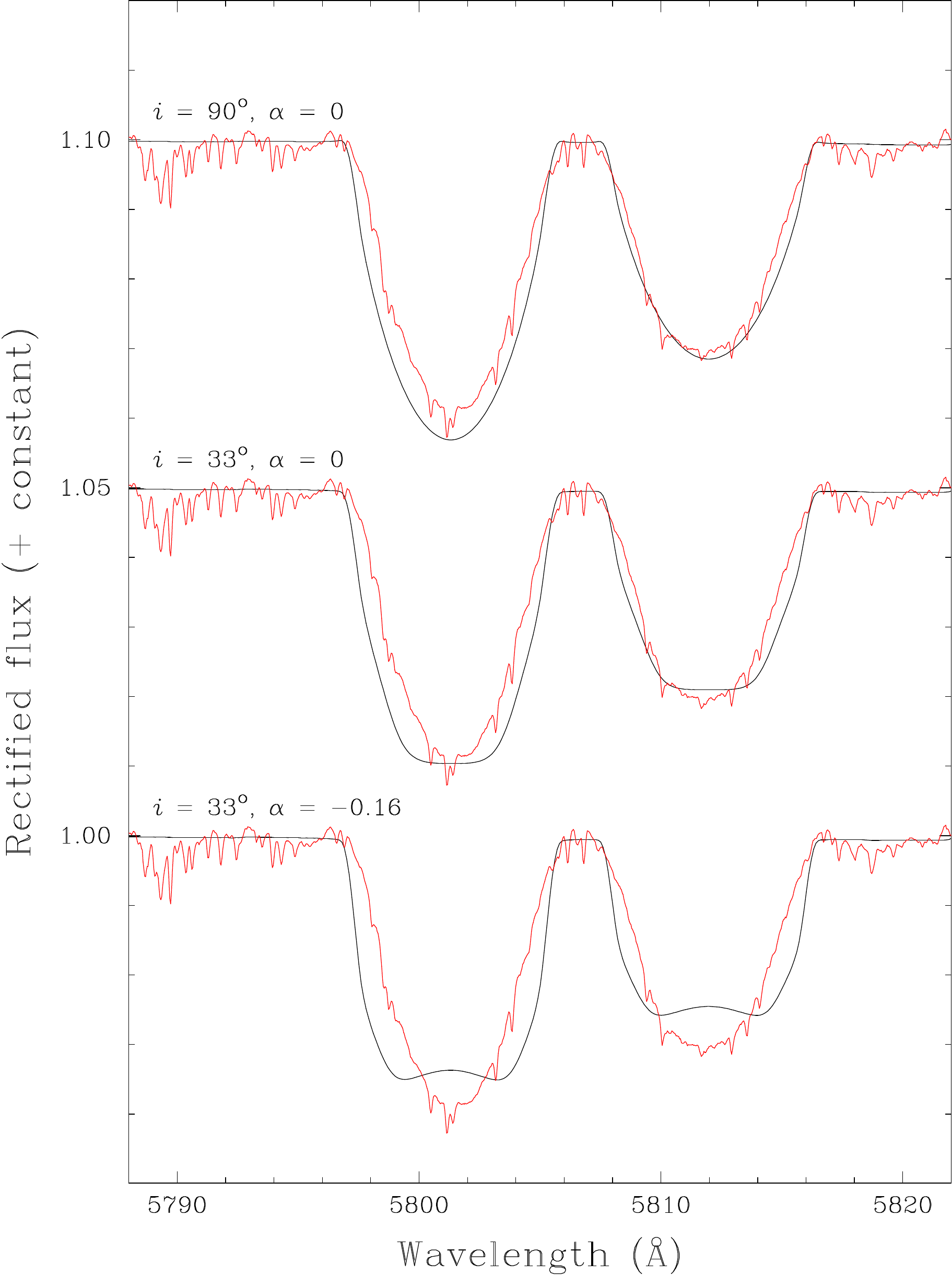}
\caption{Synthetic C\,\textsc{iv}~$\lambda\lambda$5801,~5812 spectra
  (described in Section~\ref{sec:difrot}),
  compared with the mean of observations obtained 
in 2000 December, using the
UCLES echelle spectrograph on the (then) Anglo-Australian Telescope
(Donati \& Howarth, unpublished).
The extended emission pedestal
\citep{Baschek71, Baade91} has been removed as part of the
rectification;  the numerous narrow features are the result of
telluric absorption.
}
\label{fig:figrot}
\end{figure}

\subsection{CIRs, DACs, and spots}

The phenomenological two-dimensional radiation-hydrodynamical CIR
simulations by \citeauthor{Cranmer96} (\citeyear{Cranmer96}; see also
\citealt{DavidUraz17}) employed the heuristic mechanism of
photospheric bright spots to drive locally enhanced mass outflows,
which they showed can generate DAC-like features (through velocity
plateaux, rather than directly through density enhancements).
Such spots may arise through subphotospheric convection zones
\citep{Cantiello11}, and 
subsequent observational efforts have sought to identify
corresponding spot-like surface features, and to associate
them with DAC drivers (e.g., \citealt{TahinaR14}).

$\zeta$~Pup is particularly well suited to such investigations, as it
has the longest intensive UV-spectroscopy time series of any O~star: a
16-d sequence of IUE observations in 1995 January \citep{Howarth95}.  
These reveal a DAC recurrence
timescale $T_{\rm DAC} = 19.23{\pm}0.45$~hr at that epoch.
Essentially the same timescale was found by
\citet{Reid96} in H$\alpha$, over the velocity range
$\sim{-300}:-800$\kms,
in observations obtained in 1990.

Unfortunately for the `corotating spots'
hypothesis, the \mbox{1.78-d} period is not commensurate with this DAC
timescale; in particular, the duration of the IUE time series (hence
precision of the timescale determination) rules out the possibility
that $P_{\rm phot} = 2\times T_{\rm DAC}$ with 4-$\sigma$ confidence.

\citet{TahinaR18} sought to reconcile the superficially inconsistent
timescales of the \mbox{1.78-d} photometric signal and (twice) the
$\sim$19-hr DAC recurrence period by speculating that $\zeta$~Pup may
rotate differentially.  In that case, a direct causal link between the
mooted photospheric hot-spots and CIR formation could be salvaged if
the spots occurred at higher, faster-rotating latitudes at the time of
the IUE observations than at the epochs of the space-based
photometry.\footnote{The required differential rotation is
  antisolar --  i.e., in the
  opposite sense to that observed in the Sun and solar-type stars, where the
  rotation period is shortest at the equator (e.g.,
  \citealt{Carrington1860, Benomar18}).}
This speculation is bolstered by growing evidence for differential rotation in
some stars bluewards of the granulation boundary \citep{Balona16b}.

\subsection{Differential rotation?}
\label{sec:difrot}

Direct testing of this speculation would require contemporaneous,
extensive time series of both UV spectroscopy and photometry, which
are unlikely to be available in the immediate future.  However,
differential rotation also has observable consequences for
photospheric line profiles.  To explore this, we performed
calculations using a simple prescription for latitudinally
differential angular rotation $\omega$ at colatitude $\theta$,
\begin{align*}
\frac{\omega(\theta)}{\omega_{\rm e}} &=
1 - k + k\left\{{
\frac{R(\theta)\,\sin\theta}{\reqtr} }\right\}^2
\end{align*}
(which in the spherical limit simplifies to the form commonly adopted for late-type stars,
${\omega(\theta)}/{\omega_{\rm e}} =
1 - k\cos^2\theta$).
To estimate the $k$ parameter, we suppose that at the time of the IUE
observations any spot features transited centrally (i.e., 
$\theta_{\rm s} \simeq i$),  as suggested by the large covering factor
of the DACs (i.e., the large fractional coverage of the
projected stellar disk), and that their rotation period was $2T_{\rm
  DAC}$
(with an equatorial rotation period of \pphot = 1.78~d);  these
assumptions lead to $k = -0.16$.

Three sets of model profiles are shown in Fig.~\ref{fig:figrot}, based
on general physical parameters from columns~3 and~5 of
Table~\ref{t_obs}.  We stress that these are ad hoc, ab initio model
calculations, and are not, in any sense, fits to observations (which
would properly entail exploration of a wider parameter space).
Nevertheless, the comparison with observed profiles is of some
interest; the traditional, $i \simeq 90^\circ$ ($\prot = 3.35$~d)
model provides a reasonably satisfactory match to the C\,\textsc{iv}
absorption profiles, while the $\prot = 1.78$~d model fares rather
less well.  The discrepancies with the differentially rotating model
are large enough to cast doubt on the underpinning speculation, and
hence on the proposal that DACs in $\zeta$~Pup are directly driven by
the same phenomenon that is responsible for the photometric signal.

We mention two further practical difficulties that challenge 
a model whereby DACs are the result of CIRs driven directly by
corotating photospheric hot spots:\newline
$\phantom{m}$(\textit{i}) 
It appears likely that all early-type stars with strong
  winds exhibit DACs (e.g., \citealt{Howarth89}) and yet there is currently
  little evidence for ubiquitous periodic photospheric spot activity
  (or for strictly periodic DAC activity).
In the specific case of $\zeta$~Pup, there is a clear record of
`normal' DAC behaviour in 1989 April \citep{Prinja92}, while 
contemporaneous photometry gives no indication of
significant
variability on a commensurate period (\citealt{Balona92}; $\S$\ref{hipp:phot}).\\
$\phantom{m}$(\textit{ii}) 
The DAC covering factors are large ($\gtrsim$0.5; e.g.,
\citealt{Howarth95a, Massa15}).   If we suppose that DACs were present
in $\zeta$~Pup at the epochs of the 21st-century space photometry (which, though now
untestable, seems
likely, given their \mbox{$\sim$universal} occurence), it is not obvious how
this can be reconciled  with an origin in CIRs originating from
relatively small, $\sim$equatorial spots viewed at
low axial inclination.

\subsection{Other timescales}

Part of the motivation that led \citet{TahinaR18} to seek an
association of \pphot\ with $T_{\rm DAC}$ was their discovery that the
1.78-d period is not associated exclusively with the photosphere, but
can additionally be traced in the
\HeII~$\lambda$4686 emission line
at velocities out to $\sim{\pm}400$~\kms. It is apparent, therefore,
that there is
\textit{some} connection between activity in the photosphere and 
the base of the stellar wind.

A similar association was reported by \citeauthor{Reid96}
(\citeyear{Reid96}; data obtained 1992), who found an 8.5-hour signal
in photospheric lines, with blue-to-red propagation.\footnote{This
  period was first identified by \citet{Baade86}, in data obtained in
  1984.  He proposed an interpretation in terms of sectoral-mode
  non-radial pulsations ($\ell = |m| = 2$; \citealt{Baade88a},
  \citealt{Reid96}).  The period has not been recovered in other
  datasets of comparable, or better, quality (cf.\ \citealt{Howarth14,
    TahinaR18}).  At the risk of invoking arbitrary numerology, we
  note the coincidence that $\pphot = 1.78$~d is \textit{exactly}
  5$\times$ the 8.54-hr period reported by \citeauthor{Reid96}.}  They
identified the same period in H$\alpha$ emission at
velocities more negative than $-280$~\kms, moving
\textit{red to blue} -- further evidence
for a stellar-wind signature of photospheric activity.

However, in that case the 8.5-hr signal co-existed with a 19.2-hr
signal (matching $T_{\rm DAC}$), the latter again featuring red-to-blue
migration, detected at outflow velocities from $\sim$300 to 800~\kms.  It therefore appears
possible that the low-velocity wind may respond to photospheric
drivers (of whatever nature) without that response necessarily
propagating directly to the DACs in the high-velocity wind.

\section{Summary \& conclusion}
\label{sec:summ}

We have argued that the \hipp\ astrometry for $\zeta$~Pup is reliable,
and hence that $d = 332 \pm 11$~pc ($\S$\ref{sec:hpllx}).  With this
distance, we modelled the basic physical characteristics under two
extreme assumptions:\newline (\textit{i}) $i \simeq 90^\circ$ ($\prot
\simeq 3.3$~d, \mbox{$\veq \simeq 213$~\kms,} \mbox{$\omomc \simeq
  0.6$}), and\newline (\textit{ii}) $\prot = \pphot$ (\mbox{$i \simeq
  33^\circ$}, \mbox{$\veq \simeq 390$~\kms}, \mbox{$\omomc \simeq
  0.9$}), where \mbox{\pphot = 1.78~d} is the period observed in 21st-century
broad-band, space-based photometry.

In either case, $R_* \simeq 13\rsun$, $M \simeq 25\msun$ (cf.\
Table~\ref{t_obs} and Section~\ref{sec:res1} for details).  These
characteristics are not consistent with single-star evolutionary
tracks.  As proposed by \citet{vanRensbergen96} and \citet{Vanbeveren12}, binary (or
multiple-star) interaction earlier in $\zeta$~Pup's lifetime is
probably required, and is then implicated in its runaway status, and
in its exceptionally rapid rotation.

The \mbox{1.78-d} photo\-metric variability was of smaller amplitude
(or absent) in \hipp\ photometry than more-recent datasets
($\S$\ref{hipp:phot}), confirming previous indications that secular
changes occur.

We have reviewed the possibility that the 1.78-d signal arises from
rotational modulation of photospheric starspots (i.e., that
$\pphot = \prot$).  Available modelling of the photometric data is
incapable of discriminating between hot and cool spots (or other
mechanisms, such as non-radial pulsation).  Time-series spectroscopy
resolving the line profiles should afford a reasonably clean test of
the corotating-spot hypothesis (more precisely, of the implied low
axial inclination; $\S$\ref{sec:lpv}), though existing data do not
show the expected signature.  We consider that the origin of the
\mbox{1.78-d} period remains open to question (particularly given that a separate, 2.56-d
signal was present in the \hipp\ photometry, and was similarly 
attributed to rotational modulation).

We have re-examined the putative association between \pphot, \prot,
CIRs, and DACs.    A direct causal association between DACs and corotating
interaction regions driven at \pphot\
would require an ad hoc
mechanism of differential rotation coupled to latitudinally mobile hotspots.
Exploratory line-profile calculations afford a reasonable match to
observations for $i=90^\circ$ ($\prot \simeq 3.3$~d);
lower-inclination, differentially rotating models give poorer
agreement ($\S$\ref{sec:difrot}).
We conclude that a compelling case for DACs being the result of CIRs driven
directly by corotating photospheric hotspots has yet to be made for $\zeta$~Pup.

\section*{acknowledgements}

We thank Tony Moffat, Ya\"el Naz\'e, Stan Owocki, Raman Prinja, Tahina Ramiaramanantsoa, Dominic
Reeve, Dany Vanbeveren, and Gregg Wade for useful correspondence/conversations.

\bibliographystyle{mnras}

\bibliography{IDH}

\label{lastpage}

\end{document}